\documentclass{iau} 
\usepackage{graphicx}

\title[Astronomy Data, Virtual Observatory and Education] 
{Astronomy Data, Virtual Observatory and Education}

\author[Priya Hasan \& S N Hasan ]{Priya Hasan \& S N Hasan}   

\affiliation{Maulana Azad National Urdu University, Hyderabad, India \\ email: {\tt priya.hasan@gmail.com} }

\pubyear{2021}
\volume{367}  
\jname{Education and Heritage in the era of Big Data in Astronomy}
\editors{R.M. Ros, B. Garcia, S. Gullberg, J. Moldon \& P. Rojo, eds.}

\begin{document}

\maketitle

\begin{abstract}
We shall present with examples how analysis of astronomy data can be used for an educational purpose to train users in methods of data analysis, statistics, programming skills and research problems. Special reference will be made to our IAU-OAD project `Astronomy from Archival Data' where we are in the process of building a repository of instructional videos and reading material for undergraduate and postgraduate students as well as interested participants. Virtual Observatory tools will also be discussed and applied. As this is an ongoing project, by the time of the conference we will have the projects and work done by students included in our presentation. The material produced can be freely used by the community. 
\keywords{astronomical data bases, sociology of astronomy}
\end{abstract}

\firstsection 
\section{Introduction}


Virtual learning is a learning-teaching  experience that is technology and communication enabled in a synchronous (web-conferencing) or asynchronous (self-paced) manner. Teaching is online and the teacher and learners are physically separated (in terms of place, time, or both). One of the biggest challenges to educators during the covid pandemic was virtual teaching which became the sole mode of communication. Virtual teaching has its advantages namely, convenience of access, wide global reach, inclusive content and collaborations across the globe. However, it has it's disadvantages too: essentially the lack of face-to-face interaction and the human element. The present talk describes our experiences with virtual teaching of astronomy with the help of astronomy data and Virtual Observatory tools. 

 In the early April 2020, we collaborated with Jana Vigyan Vedika (JVV) Telangana which is a member organisation of the All Indian People Science Network (AIPSN) and  a member of the National Council for Science and Technology Communication (NCSTC) in India. The NCSTC is a scientific programme of the Government of India for the popularisation of science, dissemination of scientific knowledge and inculcation of scientific temper. JVV is an effective Science Forum in the state and has a very good reach in rural and urban Telangana. 

We started off with a Six-Day Online Astronomy Course from the 18-23 May 2020 from 4:30- 6:30 pm IST for School Teachers and interested students. We conducted a quiz at the end of the course and gave certificates to participants who scored above a threshold mark. We had 475 registrations and many were given certificates with JVV.
 
In May 2020, the International Astronomical Union-Office of Astronomy for Development (IAU-OAD) issued an Extraordinary Call for Proposals that could `use astronomy in any form to help mitigate some of the impacts caused by COVID-19'. We proposed a project `Astronomy from Archival Data' which involved Educational Projects for Under-Graduate  and Post-Graduate Science Students (\cite[Possel, 2020]{Possel20}). The project trained students to use the high-quality astronomy data from various facilities. Participants were shown step-by-step techniques of accessing and analysing astronomy data from the internet, introduced to virtual observatory tools and programmimg techniques, and supported to formulate and develop projects. The training included special sessions on report writing, publishing and presenting in scientific journals\footnote{More info at: https://shristiastro.com/astronomy-from-archival-data/}. 

Almost 1000  students from 25 countries registered for the course, which was encouraging and which made it clear that students clearly require this kind of skill training. Figure \ref{fig1} shows a mosaic of (1) Our website (2) Our YouTube channel (3) A zoom session in progress (4) One of the recorded sessions.

\section{Modus Operandi}
Although virtual teaching has been there for a while, this time it required more thought and planning. With the funds we has from OAD, we first set up a website design and content. To enable interaction, we used Social Media (Facebook), Telegram as well as regular email. At regular intervals we planned Homework, Quizzes and Polls so that we could get feedback from our participants.

The project consisted of three phases. The first was made up of live zoom sessions on Saturdays and Sundays from 10-11:30 am IST with lectures and interactive hands-on sessions with us and a variety of speakers from India and abroad on Virtual Observatory Tools, Data Archives and Science Cases. All the sessions were recorded and posted online. We now have over 60 video recordings online on YouTube.
Reading material, quizzes, data sources, analysis techniques, etc. are available on our website (https://shristiastro.com/astronomy-from-archival-data/) as well as on YouTube.  The sessions ran from August to the end of November, 2020. The second phase was in December, when students could complete watching videos and select projects and interact with mentors. The third phase will be till the second half of February to give students time to work on projects of their choice. Students will present their results in the last week of February, 2021.  

Apart from us, we had a wide variety of speakers like  Mark Taylor, Luisa Rebull,  Deborah Baines, Tim Hamilton, Ajit Kembhavi, Avinash Deshpande, Sushan Konar, Kaustubh Waghmare, Eeshan Hasan etc.  to cover a wide range of junior and senior researchers. 
The talks ranged from Basics of Python, Astropy, TOPCAT (\cite[Taylor, 2003]{Taylor93}), Aladin, ESASky, Machine Learning and Coding and how to use them. We also had talks on Science Cases for projects and some hand-held sessions on projects.   

\begin{figure}[h]
\begin{center}
\includegraphics[scale=0.5]{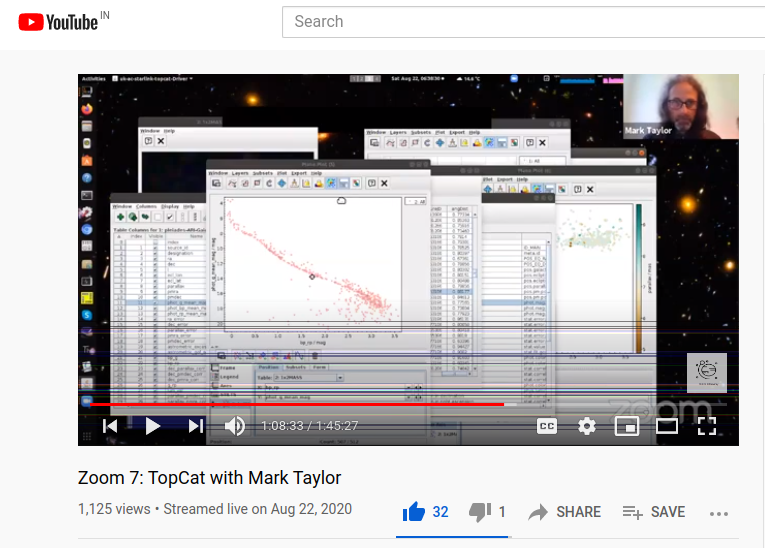} 
\caption{A Zoom Session in progress TOPCAT with Mark Taylor}
\label{fig1}
\end{center} 
\end{figure}

Figure \ref{fig1} shows a Zoom Session in progress: TOPCAT with Mark Taylor. 

\section{Projects}
Students were free to opt for one or more projects of their choice. Mentors have been having sessions with their groups to individually address issues and work on the projects. The list of projects offered are as follows: 
\begin{itemize}
\item 
Exoplanets (Tim Hamilton/Priya Hasan)

This project involves analysis of Kepler light curves to derive parameters for Exoplanets including distance to the star, mass, radius and hence density. This and the next three projects were explained in a step-by-step procedure  by Tim Hamilton and is available on our YouTube website. Enthusiastic participants are encouraged to try their skills with Transiting Exoplanet Survey Satellite (TESS) data.  
\item 
Globular Clusters Photometry (Tim Hamilton/Priya Hasan)

This is an exercise in doing Aperture Photometry on images of globular clusters from the SDSS data. Color-magnitude diagrams for the globular cluster are constructed. 
\item 
Globular Clusters Color-Magnitude Diagrams: Finding the ages and distances of Milky Way globular clusters (Tim Hamilton/Priya Hasan)

For a sample 11 globular clusters, participants had to plot the color-magnitude diagram showing the cluster data and the best-fitting isochrone together.
\item  
Gravitational Lensing (Tim Hamilton/Priya Hasan)

For a given list, extracting close-in image of each gravitational lens system, showing the lens and the Einstein Ring is to be done. Students need to calculate the total mass,  stellar mass,  mass of dark matter and  dark matter as a fraction of total mass.
\item 
The Radio Pulsar-Magnetar Connection (Sushan Konar) 

This project involved a comparison of the intrinsic properties of radio-pulsars and magnetars using available data. 
\item 
Star clusters with Gaia (TOPCAT) (Priya Hasan)

Extracting Gaia data and plotting the color-magnitude diagram using TOPCAT. 
\item 
Python and Gaia

Using python code with Gaia data (Priya Hasan) 
\item
The spatial distribution of pulsars and the spiral structure of the galaxy (Avinash Deshpande)
\item 
Galaxy Morphology (S N Hasan) 

Studying galaxy morphology using decomposition of luminosity profiles  using GALFIT (\cite[Peng, 2010]{Peng10})
\item 
Others

Participants were free to propose their own projects based on the material covered. 
\end{itemize}

 We hope that some of these students may even use this for applications in PhD programs/Graduate school as well as a start to serious research careers. 
    
\section{Lessons Learnt}

This project is close to completion. The Internal Virtual Observatory Alliance (IVOA) supports this project. Participants will present their projects and their presentations will be added to the repository \footnote{As a team, we run Shristi Astronomy (https://shristiastro.com/). Details of our activities are on the website and the reader is encouraged to have  a look. All the sessions have been recorded and are available on youtube https://www.youtube.com/c/shristiastronomy}.

In the course of our sessions we learned a few lessons which we would like to share.
\begin{itemize}
\item
Programs like this are very useful to students since it shows them important steps in data analysis and techniques which are generally never taught as part of a course. To get the true flavour of a research career, it is very useful if students try their hand on some projects. 
\item
Not all who register participate and hence there has to be constant monitoring of the participants. Interaction via social media, email, etc has to be maintained to keep in tune with issues faced by participants.  This can be very time consuming as well as difficult.
\item
Planning of the talks has to be done carefully since most potential participants had their own online sessions on. We selected weekends since most students/participants do not have classes then. As time progressed, classes as well as examinations were on which made it difficult for students/participants to attend our classes. 

\item There are various issues like internet connectivity problems, power as well as overlapping sessions. Hence all our activities were recorded and made available on YouTube  so that participants could attend at their convenience. 

\item A few of the participants were from non-English speaking countries, hence we enabled subtitles on our videos for their benefit. 

\item
It is good to have a variety of speakers ranging from students, research scholars, post-doctoral fellows and junior and senior faculty since it can match the level of the participant and it makes the environment more inclusive. 

\item
While sessions were on, gradually many students had regular classes and exams due to which the numbers dropped. Since the repository we created is available online, we hope that students can access the material at their convenience.
 
\end{itemize}

We have received approval for a new IAU-OAD project `AstroSprint' which will be a follow-up of this project in the form of three workshops in the year 2021. The first will still be online and we hope the next two in July and November 2021 wgich will be offline where we can have interactions in-person. We think this was a very timely and purposeful initiative and it was a very good learning exercise. The support of the IAU-OAD is appreciated. We hope to continue with this work of developing a repository of instructive videos and study material that can be used for astronomy research.

\end{document}